\documentclass[aps,11pt,prd,groupedaddress,nofootinbib,notitlepage,eqsecnum,preprintnumbers]{revtex4-2}

\usepackage[utf8]{inputenc}
\usepackage{graphicx}
\usepackage{bm}
\usepackage{amsmath, amssymb}
\usepackage{mathtools}
\usepackage{physics}
\usepackage[english]{babel}
\usepackage[colorlinks=true, allcolors=blue]{hyperref}
\usepackage{url}
\usepackage{hyperref}
\usepackage{cleveref}
\usepackage[dvipsnames,svgnames]{xcolor}
\usepackage[normalem]{ulem}
\usepackage{soul}
\usepackage{bbold}
\usepackage[lofdepth,lotdepth,caption=false]{subfig}
\usepackage{lipsum}
\usepackage{mathrsfs}
\usepackage{stmaryrd}
\usepackage{yhmath}
\usepackage{comment}
\usepackage{tikz-feynman}
\usepackage{subfig}

\DeclareMathOperator{\arccosh}{arccosh}
\DeclareMathOperator{\arcsinh}{arcsinh}
\DeclareMathOperator{\Erf}{erf}

\newcommand{\bk}{\bm{k}}
\newcommand{\bp}{\bm{p}}

\newcommand{\bx}{\bm{x}}
\newcommand{\bl}{\bm{l}}
\newcommand{\bq}{\bm{q}}
\newcommand{\mpl}{M_{\text{Pl}}}
\newcommand{\be}{\bm{e}}

\definecolor{elpurple}{HTML}{a300ff}

\begin{document}

\title{Unpolarized Low-Frequency Tail of Causal Chiral Gravitational Waves}

\author{\textsc{Tikinas Chabour$^{a}$}}
    \email{{tikinas.chabour}@{ens.psl.eu}}
\author{\textsc{Guillem Domènech$^{a,b}$}}
    \email{{guillem.domenech}@{itp.uni-hannover.de}}
\author{\textsc{Alexander Ganz$^{a}$}}
    \email{{alexander.ganz}@{itp.uni-hannover.de}}
\author{\textsc{Jan Tränkle$^{a}$}}
    \email{{jan.traenkle}@{itp.uni-hannover.de}}
    
\affiliation{$^a$Institute for Theoretical Physics, Leibniz University Hannover, Appelstraße 2, 30167 Hannover, Germany}
\affiliation{$^b$Max Planck Institute for Gravitational Physics,
Albert Einstein Institute, 30167 Hannover, Germany}

\begin{abstract}
    Parity violation in the early Universe may lead to a chiral gravitational wave background. We show that, for finite-time parity-violating sources, the low-frequency tail of the generated chiral gravitational wave background is universally unpolarized, and that the chirality fraction decreases linearly with frequency. Our result generally applies to any parity-violating gravitational-wave source that appears quadratically in the wave equation. We consider both Gaussian sources and explicit trispectrum templates. We attribute the lack of chirality of the low-frequency tail to the fact that, on large scales, gravitational wave generation is almost equally inefficient for both helicities.
\end{abstract}

\maketitle

\section{Introduction} \label{introduction}

The detection and characterization of a stochastic gravitational-wave background (GWB) of cosmological origin would open a unique window into the violent processes in the early Universe that could have sourced such a signal and would also probe new physics beyond the standard (cosmological) model.
The tentative detection of a GWB signal at nano-Hertz frequencies by pulsar timing arrays (PTAs) has sparked renewed interest in cosmological scenarios capable of producing such loud GWs \cite{NANOGrav:2023hvm}, including first-order phase transitions \cite{Witten:1984rs}, curvaton scenarios \cite{Bartolo:2007vp}, or enhanced primordial perturbations \cite{Domenech:2021ztg,Chang:2022vlv}.
Interestingly, potential polarization of the GWB carries information about parity-violating processes in the early Universe \cite{Komatsu:2022nvu}, which could be probed with upcoming GW detectors such as the Einstein Telescope \cite{Caporali:2026qhc}, LISA-Taiji \cite{Orlando:2020oko}, and PTAs \cite{Xu:2026ltq}; see also Refs.~\cite{Seto:2008sr,Nishizawa:2009jh} for earlier works.

While the generation of parity violation in standard inflationary scenarios is restricted by no-go theorems \cite{Cabass:2022rhr}, the fundamental requirement of $\rm CP$ violation for successful baryogenesis strongly motivates the study of theories beyond the simplest models.
In fact, there exists a large number of parity-violating models that could produce a circularly polarized, that is, chiral, GWB. These include, for instance, general axion and dark photon models \cite{Machado:2018nqk,Machado:2019xuc,Salehian:2020dsf,Ratzinger:2020oct,Banerjee:2021oeu,Garcia-Bellido:2023ser}, ghost inflation \cite{Arkani-Hamed:2003juy,Garcia-Saenz:2023zue}, scalar-tensor induced GWs \cite{Bari:2023rcw}, thermalized neutrinos \cite{Gubler:2022zmf}, and helical magnetic fields \cite{Okano:2020uyr,Ragavendra:2026fgs}.
Furthermore, parity-violating modifications of the gravitational sector, e.g.~through a Chern-Simons term \cite{PhysRevD.68.104012,Creque-Sarbinowski:2023wmb} or in teleparallel gravity models \cite{Iosifidis:2018zwo,Li:2022vtn}, can source a chiral (scalar-induced) GWB \cite{Zhang:2022xmm,Feng:2023veu, Zhang:2023scq,Zhang:2025mps,Zhang:2026dic}.
An upper limit on the amplitude of a chiral cosmological GWB produced before the electroweak epoch was derived in \cite{Gorji:2026tln}. In the aforementioned models, one typically finds an unpolarized low-frequency tail of the GW spectrum.

However, to the best of our knowledge, no general explanation for such unpolarization has been provided so far. In this paper, we show that an unpolarized low-frequency, or equivalently infrared (IR), tail is expected on general grounds for finite-time, parity-violating sources. Our work builds upon Ref.~\cite{Cai:2019cdl}, which showed that unpolarized, finite-time GW sources always yield a GW spectrum with a universal $f^3$ IR tail, which follows from causality reasons -- namely, that the energy momentum tensor on super-Hubble scales must be constant. We find a universal linear scaling law for the chirality fraction.

Establishing a universal behavior is crucial for observational prospects. For example, if the peak of cosmic GW spectra lies at frequencies too high for current and near-future GW detectors, the IR tail often represents the only accessible window into these parity-violating processes. Moreover, demonstrating a universal scaling for the chirality fraction helps reliably distinguish genuine primordial signals from astrophysical foregrounds.

The rest of the paper is structured as follows. 
We introduce our general formalism in Sec.~\ref{sec:ir-vector}. Although we focus mostly on vector-induced GWs, the formalism is general. In Sec.~\ref{sec:ir-gaussian}, we focus on Gaussian polarized sources and show that the resulting vector-induced GWs have a chirality fraction that scales linearly with frequency in the IR tail. As examples, we consider a Dirac-delta and a log-normal primordial spectrum. In Sec.~\ref{sec:Trispectrum}, we turn our attention to scalar-induced GWs from a parity-violating primordial trispectrum. For concreteness, we work with two templates for the trispectrum and demonstrate that the chirality parameter also scales linearly in frequency in the IR tail. We conclude our work in Sec.~\ref{sec:summary}. Details of the calculations are provided in the Appendices.

\section{Parity violation in cosmic GWs: General framework} \label{sec:ir-vector}

In this section, we present the general framework for subsequent calculations of quadratic, parity-violating sources. To do so, we consider transverse-traceless tensor perturbations $h_{ij}$ on a flat Friedmann-Lemaître-Robertson-Walker (FLRW) background,
\begin{equation}
    ds^2=a(\eta)^2\left(-\mathrm{d}\eta^2+\left(\delta_{ij}+h_{ij}\right)\mathrm{d}x^i\mathrm{d}x^j\right) \,,
\end{equation}
where $a$ is the scale factor, $\eta$ denotes conformal time and we neglected scalar and vector perturbations without loss of generality.
In Fourier space, we express tensor modes as
\begin{align}\label{eq:hijfourier}
    h_{ij}(\eta,\mathbf{x})=\sum_\lambda\int \frac{d^3\bk}{(2\pi)^{3/2}}\,e^\lambda_{ij}(\hat{\bk})h_{\bk,\lambda}(\eta)e^{i\bk\cdot\bx} \,,
\end{align}
where $\hat{\bk}=\mathbf{{k}}/k$ with $k=|\mathbf{{k}}|$, and $e^\lambda_{ij}(\hat{\bk})$ are the polarization tensors, explicitly given in App.~\ref{app:pol-basis}. From now on, we use circular polarization for convenience and denote the right and left polarizations by $\lambda=\pm$, respectively. Note that we use normalized polarization tensors, that is, $e^\lambda_{ij}(\hat{\bk})e^{\lambda'*}_{ij}(\hat{\bk})=\delta^{\lambda\lambda'}$ with an implicit sum over repeated indices and an asterisk denoting complex conjugation. The reality condition yields $e^{\lambda'*}_{ij}(\hat{\bk})=e^\lambda_{ij}(-\hat{\bk})=e^{-\lambda}_{ij}(\hat{\bk})$, as well as $h^*_{\bk,\lambda}(\eta)=h_{-\bk,\lambda}(\eta)=h_{\bk,-\lambda}(\eta)$. Also note that since the spatial metric is flat, that is $\delta_{ij}$, we do not make a distinction between upper and lower spatial indices. The same applies to $\lambda$.

The mode functions $h_{\bk,\lambda}(\eta)$ in Eq.~\eqref{eq:hijfourier} satisfy the standard GW equation in Fourier space, see, e.g., Ref.~\cite{Cai:2019cdl}, namely
\begin{equation}\label{eq:tensor-eom}
    {h}''_{\bk,\lambda}+2\mathcal{H}h_{\bk,\lambda}'+k^2h_{\bk,\lambda}=2 \mpl^{-2} a^2{\cal S}_{\bk,\lambda}(\eta) \,,
\end{equation}
where $\mpl^{-2}=8\pi G$ is the reduced Planck mass and ${\cal S}_{\lambda,\bk}(\eta)$ is a general source in Fourier space stemming from the matter content, which we specify below in Sec.~\ref{sec:mattercontent}. In Eq.~\eqref{eq:tensor-eom}, $\mathcal{H}\coloneqq a'/a$ is the so-called conformal Hubble parameter and a prime denotes differentiation with respect to $\eta$, that is $a'=da/d\eta$. Formally, Eq.~\eqref{eq:tensor-eom} can be solved using the Green's function method, which yields
\begin{align}\label{eq:greensolution}
h_{\bk,\lambda}(\eta)=\frac{2}{a(\eta)\mpl^2}\int_0^\infty d\tilde \eta \,a^3(\tilde \eta) G(\eta,\tilde \eta){\cal S}_{\bk,\lambda}(\tilde\eta)\,,
\end{align}
where the tensor modes' retarded Green function is given by \cite{Cai:2019cdl}
\begin{align}\label{eq:greens}
    G(\eta,\tilde{\eta})=\frac{1}{k}\sin \left(k(\eta-\tilde{\eta})\right)\Theta(\eta-\tilde{\eta}) \,.
\end{align}

Since we are interested in the GWB today, all frequencies of interest are sub-Hubble, that is, we are interested in the late time limit $k\eta\gg1$, or equivalently $f\gg {\cal H}$. On such deep sub-Hubble scales tensor perturbations behave as free GWs, and one can define an effective energy density by integrating out the high-frequency GW modes through a suitable space-time average \cite{Isaacson:1968hbi,Isaacson:1968zza,Domenech:2021ztg}. Using the so-called Isaacson prescription, the energy density of GWs per polarization and per logarithmic wavenumber is given by \cite{Cai:2019cdl}
\begin{equation}\label{eq:rhoGW}
    \rho^\lambda_\mathrm{GW}
    =\frac{\mpl^2}{8 a^2}\frac{k^3}{2\pi^2}
    \langle {h}'_{\bk,\lambda}(\eta){h}_{\bk,\lambda}^{'*}(\eta)+k^2h_{\bk,\lambda}(\eta)h^{*}_{\bk,\lambda}(\eta)\rangle^! \,,
\end{equation}
where the square brackets denote a space-time average and, via the ergodic theorem, also an ensemble average. Note that in deriving Eq.~\eqref{eq:rhoGW} we assumed that
\begin{align}\label{eq:assumption}
\langle {h}_{\bk,\lambda}(\eta){h}_{\bq,\lambda'}^{*}(\eta)\rangle=\delta_{\lambda\lambda'}\delta^{3}(\bk -\bq) \langle {h}_{\bk,\lambda}(\eta){h}_{\bk,\lambda}^{*}(\eta)\rangle^!\,,
\end{align}
where the exclamation mark emphasizes that the Kronecker and Dirac delta have been factored out. Note that the appearance of $\delta^{3}(\bk -\bq)$ follows from the homogeneity of the background. For the models we consider, circular polarizations are independent and, therefore, we have $\delta_{\lambda\lambda'}$.

Using the Green's method solution Eq.~\eqref{eq:greensolution}, the resulting GW energy density parameter, that is $\Omega^\lambda_\mathrm{GW}=\rho^\lambda_{\rm GW}/(3H^2M_{\rm pl}^2)$ with $H={\cal H}/a$,  then reads
\begin{align}\label{eq:omegaGW}
    \begin{split}
        \Omega^\lambda_\mathrm{GW}(k) = & \, \frac{k^3}{12\pi^2 a^2{\cal H}^2 \mpl^4}\int_{0}^{\eta} d\eta_1\int_{0}^{\eta} d\eta_2 \,
        a^3 (\eta_1)a^3 (\eta_2)\cos \left(k(\eta_1-\eta_2)\right) \langle{\cal S}_{\bk,\lambda}(\eta_1){\cal S}^*_{\bk,\lambda}(\eta_2)\rangle^!\,,
    \end{split}
\end{align}
where we used that on sub-Hubble scales $h'_{\bk,\lambda}\approx k h_{\bk,\lambda}$. To simplify the notation, we do not write the time dependence of the source explicitly in what follows. Note that if $\langle{\cal S}_{\bk,\lambda}{\cal S}^*_{\bk,\lambda}\rangle^!$ has no $k$ dependence then Eq.~\eqref{eq:omegaGW} directly yields $\Omega^\lambda_\mathrm{GW}\propto k^3$ regardless of polarization. This is, naively, always the case for GWs with frequencies much smaller than the source time scale of GW generation \cite{Cai:2019cdl}. Note that, although eventually correct at leading order, this argument neglects contributions from the polarization projection factors. As we shall see, the projection factors are crucial when computing the chirality parameter, namely \cite{Orlando:2020oko,Cai:2021uup}
\begin{equation}\label{eq:pi-def}
    \Pi_\text{GW} \coloneqq  \frac{\Omega_\text{GW}^+-\Omega_\text{GW}^{-}}{\Omega_\text{GW}^+ + \Omega_\text{GW}^{-}}\,.
\end{equation}

\subsection{Matter content\label{sec:mattercontent}}

For the matter content, we consider a quadratic energy-momentum tensor with contributions from a vector $A_i$ and a scalar $\phi$, which can be generally written as \cite{Cai:2019cdl}
\begin{align}\label{eq:tijexamples}
    T^A_{ij}(\eta,\bx)=A_{i}(\eta,\bx)A_{j}(\eta,\bx) \quad{\rm and}\quad T^\phi_{ij}(\eta,\bx)=\partial_i\phi(\eta,\bx)\partial_j\phi(\eta,\bx)\,.
\end{align}
We consider the vector field to be transverse and treat it separately from the scalar for simplicity.\footnote{We could also consider scalar-tensor interactions \cite{Bari:2023rcw}, where one effectively has $T_{ij}=4\phi\Delta h_{ij}+4\phi'h_{ij}'$. Our results also apply to this case.} For consistency, we decompose the vector field in the helicity basis and in Fourier space as
\begin{align}
A_i=\sum_s\int \frac{d^3\bk}{(2\pi)^{3/2}}\,e^s_{i}(\hat{\bk})A_{\bk,s}(\eta)e^{i\bk\cdot\bx} \,,
\end{align}
where we use $s=\pm$ for the polarizations to avoid reusing $\lambda$, which we use exclusively for the GWs. The reality condition imposes $e^{s*}_{i}(\hat{\bk})=e^s_{i}(-\hat{\bk})=e^{-s}_{i}(\hat{\bk})$ and $A^*_{\bk,s}(\eta)=A_{-\bk,s}(\eta)=A_{\bk,-s}(\eta)$.
We expand the scalar field, which has no polarization, in a similar way. From now on, we drop the hats of the wavenumber inside the polarization tensors to avoid cluttered notation.
From Eq.~\eqref{eq:tijexamples} it follows that the source term in Eq.~\eqref{eq:tensor-eom} reads
\begin{align}\label{eq:vectorsource}
        {\cal S}^{A}_{\bk,\lambda}(\eta)&=\sum_{s_1,s_2}\int\frac{d^3q}{(2\pi)^{3/2}}e^{\lambda*}_{ij}({\bk})e^{s_1}_{i}({\mathbf{\bq}})e^{s_2}_{j}({\bk-\bq})A_{\bq,s_1}(\eta)A_{\bk-\bq,s_2}(\eta)\,,\\
       {\cal S}^{\phi}_{\bk,\lambda}(\eta)&=\int\frac{d^3q}{(2\pi)^{3/2}}e^{\lambda*}_{ij}({\bk}){q}_i q_j\,\phi_{\bq}(\eta)\phi_{\bk-\bq}(\eta)\,,
\end{align}
for the vector and scalar parts, respectively.
For the remainder of the section, we focus on the vector case only. The scalar field $\phi$ can be absorbed in $A_i$ by including a longitudinal mode in the vector, or simply replacing $e^{s_1}_{i}({\mathbf{\bq}})\to i\,q_i$ and dropping spin indices. We discuss the scalar in detail in Sec.~\ref{sec:Trispectrum}.

For the vector source \eqref{eq:vectorsource}, the two point function in Eq.~\eqref{eq:omegaGW} is given by\footnote{If we consider the general correlation function $\langle{\cal S}^A_{\bk,\lambda}{\cal S}^{A*}_{\bk,\lambda'}\rangle$ we find that the terms where $\lambda\neq \lambda'$ yields a dependence on the sum of azimuthal angles which vanishes after integration. Thus, we consider only $\lambda=\lambda'$, consistent with our assumption in Eq.~\eqref{eq:assumption}.}
\begin{align}\label{eq:sSSS}
\langle{\cal S}^A_{\bk,\lambda}{\cal S}^{A*}_{\bk,\lambda}\rangle^!=\sum_{s_1,s_2,s_3,s_4}\int\frac{d^3qd^3p}{(2\pi)^{3}}{\cal E}^\lambda_{s_1s_2s_3s_4}(\bk,\bq,\bp)\langle A_{\bq,s_1}A_{\bk-\bq,s_2}A^*_{\bp,s_3}A^*_{\bk-\bp,s_4}\rangle^!\,,
\end{align}
where we defined
\begin{align}\label{eq:calEE}
{\cal E}^\lambda_{s_1s_2s_3s_4}(\bk,\bq,\bp)\coloneqq e^{\lambda*}_{ij}({\bk})e^{s_1}_{i}({\mathbf{\bq}})e^{s_2}_{j}({\bk-\bq})e^{\lambda}_{ab}({\bk})e^{s_3*}_{a}({\bp})e^{s_4*}_{b}({\bk-\bp})\,.
\end{align}
A useful explicit expression to evaluate ${\cal E}^\lambda_{s_1s_2s_3s_4}(\bk,\bq,\bp)$ is that, for arbitrary vectors $\bk$, $\bq$ and $\bl$,
\begin{align}\label{eq:TheQs}
Q^\lambda_{s_1s_2}(\bk,\bq,\bl)\coloneqq e^{\lambda*}_{ij}({\bk})e^{s_1}_{i}({\mathbf{\bq}})e^{s_2}_{j}({\bl})=\frac{1}{4} s_1s_2 e^{-i \lambda  (\varphi_q+\varphi_l)} (1+\lambda s_1\cos\theta_q) (1+\lambda s_2 \cos \theta_l)\,,
\end{align}
where $\theta_{q/l}$ and $\varphi_{q/l}$ are the polar and azimuthal angles of $\bq$ and $\bl$ with respect to $\bk$. When $\bl=\bk-\bq$ one has that $\varphi_l=\varphi_q$ and $\cos\theta_l=(k-q\cos\theta_q)/|\bk-\bq|$. It is then clear that we can readily evaluate Eq.~\eqref{eq:calEE} as\footnote{For the scalar-tensor interaction of Ref.~\cite{Bari:2023rcw} one has instead $Q^\lambda_{\lambda'}(\bk,\bl)=e^{\lambda*}_{ij}({\bk})e^{\lambda'}_{ij}({\bl})=\frac{1}{4} e^{-2 i\lambda\varphi_l} (1+\lambda\lambda'
   \cos\theta_l)^2$ where $\bl$ is the momentum of the first-order tensor mode. It then follows that Eq.~\eqref{eq:sSSS} for the scalar-tensor case becomes
\begin{align}\label{eq:scalartensor}
\langle{\cal S}^A_{\bk,\lambda}{\cal S}^{A*}_{\bk,\lambda}\rangle^!=16\sum_{\lambda'}\int\frac{d^3q}{(2\pi)^{3}}\big|Q^\lambda_{\lambda'}(\bk,\bk-\bq)\big|^2\left(\langle h'_{\bk-\bq,\lambda'}h^{\prime *}_{\bk-\bq,\lambda'}\rangle^!\langle \phi'_{\bq}\phi^{\prime *}_{\bq}\rangle^!-|\bk-\bq|^2\langle h_{\bk-\bq,\lambda'}h^*_{\bk-\bq,\lambda'}\rangle^!\langle \phi_{\bq}\phi^*_{\bq}\rangle^!\right)\,.
\end{align}}
\begin{align}\label{eq:EforQs}
{\cal E}^\lambda_{s_1s_2s_3s_4}(\bk,\bq,\bp)=Q^\lambda_{s_1s_2}(\bk,\bq,\bk-\bq)Q^{\lambda*}_{s_3s_4}(\bk,\bp,\bk-\bp)\,,
\end{align}
where we considered only the $\phi\Delta h_{ij}$ term for simplicity.

One may further decompose the four-point correlation into products of two-point functions and its connected part, as in Ref.~\cite{Bartolo:2004if}. That is, in general, we can write
\begin{equation}\label{eq:four-point-a}
    \langle A^{\bq}_{s_1} A^{\bk-\bq}_{s_2}A^{*\bl}_{s_3} A^{*\bp-\bl}_{s_4}\rangle = \langle A^{\bq}_{s_1} A^{*\bl}_{s_3}\rangle \langle A^{\bk-\bq}_{s_2} A^{*\bp-\bl}_{s_4}\rangle + \langle A^{\bq}_{s_1} A^{*\bp-\bl}_{s_4}\rangle \langle A^{\bk-\bq}_{s_2} A^{*\bl}_{s_3}\rangle + \langle A^{\bq}_{s_1} A^{\bk-\bq}_{s_2}A^{*\bl}_{s_3} A^{*\bp-\bl}_{s_4}\rangle_{\mathrm{c}} \,,
\end{equation}
where the connected part -- denoted with a ``c'' subscript -- vanishes for Gaussian fluctuations. Unfortunately, we are not aware of a general template for the connected four-point function in the $k\to 0$ limit and for general spins. It is, nevertheless, interesting to note that the IR tail of the GW spectrum comes from the collapsed limit of the trispectrum. Namely, if we consider $\langle A_{\bk_1,s_1}A_{\bk_2,s_2}A_{\bk_3,s_3}A_{\bk_4,s_4}\rangle$ with $\bk_1+\bk_2+\bk_3+\bk_4=0$, the source of the IR tail \eqref{eq:sSSS} comes from the limit $\bk_1+\bk_2=\bk_3+\bk_4=0$, with $\bk_1=\bq$ and $\bk_4=\bp$. In the next sections, we first consider Gaussian fluctuations of a vector field and then the connected part of the trispectrum for a scalar field, for analytical viability.

\section{IR tail from parity-violating Gaussian sources} \label{sec:ir-gaussian}

We now consider that the vector field is Gaussian and that circular polarizations are, for simplicity, uncorrelated. This means that the two-point correlation function of the vector fluctuations is given by
\begin{align}\label{eq:assumptionA}
\langle {A}_{\bk,s}(\eta){A}_{\bk',s'}^{*}(\eta)\rangle=\delta_{ss'}\,\delta^{3}(\bk -\bk') P_{s}(k,\eta)\,,
\end{align}
where $P_{s}(k,\eta)$ is the power spectrum for each polarization of the vector field $A_i$. From Eqs.~\eqref{eq:sSSS} and \eqref{eq:four-point-a} we then see that
\begin{align}
\langle{\cal S}^A_{\bk,\lambda}{\cal S}^{A*}_{\bk,\lambda}\rangle^!=\sum_{s_1,s_2,s_3,s_4}\int&\frac{d^3q}{(2\pi)^3}  P_{s_1}(q)P_{s_2}(|\bk-\bq|)\nonumber\\&\times\left({\cal E}^\lambda_{s_1s_2s_3s_4}(\bk,\bq,\bq)\delta_{s_1s_3}\delta_{s_2s_4}+{\cal E}^\lambda_{s_1s_2s_3s_4}(\bk,\bq,\bk-\bq)\delta_{s_1s_4}\delta_{s_2s_3}\right)\,,
\end{align}
where we keep the time dependence implicit. Further noting that ${\cal E}^\lambda_{s_1s_2s_3s_4}(\bk,\bq,\bk-\bq)={\cal E}^\lambda_{s_1s_2s_4s_3}(\bk,\bq,\bq)$ we may simply write
\begin{align}\label{eq:sourcegaussian}
\langle{\cal S}^A_{\bk,\lambda}{\cal S}^{A*}_{\bk,\lambda}\rangle^!=2\sum_{s_1,s_2}\int&\frac{d^3q}{(2\pi)^3}  \big|Q^\lambda_{s_1s_2}(\bk,\bq,\bk-\bq)\big|^2\, P_{s_1}(q)P_{s_2}(|\bk-\bq|)\,,
\end{align}
where we already summed over $s_3$ and $s_4$ and used Eq.~\eqref{eq:EforQs} to write ${\cal E}^\lambda_{s_1s_2s_3s_4}$ in terms of $Q^\lambda_{s_1s_2}$. 

Let us now investigate the IR limit of the resulting GW spectrum. We first study the general case along the lines of Ref.~\cite{Cai:2019cdl} and later turn to two particular examples of the power spectrum that can be computed analytically.

\subsection{The universal IR scaling}

In the general case, we define the IR limit as those frequencies, or wavenumbers, which are much smaller than any scale in the system \cite{Cai:2019cdl}. On one hand, if the source is active for a finite amount of time, say $\Delta\eta_*=\eta_{\rm end}-\eta_{\rm start}\approx \eta_{\rm end}$, then we are interested in the $k\Delta\eta_*\ll 1$ regime. Note that there is also a typical associated wavenumber $q_*=\eta_{\rm end}^{-1}\approx \Delta\eta^{-1}_*$, that is the comoving Hubble scale at the end of the sourcing, below which the contribution to the GW spectrum is negligible because the scalar/vector fluctuations are still super-Hubble. On the other hand, if there is a peak in the power spectrum, say at $q_*$, then we require $k\ll q_*$. Note that, for simplicity, we purposely use the same notation $q_*$ to denote the typical scale of the system. In those cases, we may simply look at the $k\ll q$ limit of Eq.~\eqref{eq:sourcegaussian}. We also evaluate the GW spectrum at some point during the radiation-dominated era, well after the generated GWs propagate as free waves. 

We start by looking at the $k\ll q$ limit of the projection factors in Eq.~\eqref{eq:sourcegaussian}. From Eq.~\eqref{eq:TheQs}, we find, after Taylor expansion, that
\begin{align}\label{eq:Qexpansion}
\big|Q^\lambda_{s_1s_2}(k\ll q)\big|^2\approx \frac{1}{16} (1+\lambda s_1\cos\theta_q)^2 \left((1-\lambda  s_2 \cos\theta_q)^2+2\frac{k}{q}  \sin ^2\theta_q(1-\lambda  s_2 \cos\theta_q)\right)+{\cal O}(k^2/q^2)\,.
\end{align}
Since we are interested only in the leading terms, we may neglect the $\bk$ dependence in $P_{s_2}(\bk-\bq)$. We checked explicitly that the expansion in $P_{s_2}(\bk-\bq)$ does not matter for our purposes. In that limit, Eq.~\eqref{eq:sourcegaussian} explicitly reads
\begin{align}\label{eq:sourcegaussian2}
\langle{\cal S}^A_{\bk,\lambda}{\cal S}^{A*}_{\bk,\lambda}\rangle^!=4\pi\int&dq \, q^2\int_{-1}^{1} d\cos\theta_q \left(\big|Q^\lambda_{++}\big|^2\, P_{+}^2(q)+\big|Q^\lambda_{--}\big|^2\, P_{-}^2(q)+2\big|Q^\lambda_{+-}\big|^2\, P_{+}(q)P_{-}(q)\right)\,,
\end{align}
where we already performed the integration over the azimuthal angle. Using Eq.~\eqref{eq:Qexpansion} and integrating over $\theta_q$, we find that
\begin{align}\label{eq:SASUM}
\langle{\cal S}^A_{\bk,+}{\cal S}^{A*}_{\bk,+}\rangle^!+\langle{\cal S}^A_{\bk,-}{\cal S}^{A*}_{\bk,-}\rangle^!\approx\frac{8\pi}{15}\int&dq \, q^2\left( P_{+}^2(q)+ P_{-}^2(q)+12P_{+}(q)P_{-}(q)\right)\,,
\end{align}
while
\begin{align}\label{eq:SADIFF}
\langle{\cal S}^A_{\bk,+}{\cal S}^{A*}_{\bk,+}\rangle^!-\langle{\cal S}^A_{\bk,-}{\cal S}^{A*}_{\bk,-}\rangle^!\approx\frac{16\pi}{15}k\int&dq \, q\left( P_{+}^2(q)- P_{-}^2(q)\right)\,.
\end{align}
Note how the difference in GW polarization power starts linear in $k$ at leading order, while the sum of power is $k$-independent. For the latter, we conclude that the GW spectrum for each polarization scales as $k^3$ in the IR tail, as in Ref.~\cite{Cai:2019cdl}. Thus, we may say that the causal IR tail of the chiral GW spectrum is unpolarized.

We may still call it chiral GWB, though, because there is a small chirality fraction. From Eqs.~\eqref{eq:SASUM} and \eqref{eq:SADIFF}, we find that the chirality faction \eqref{eq:pi-def} is given by
\begin{align}
\Pi_{\rm GW}(k\ll q_*)\sim 2\frac{k}{q_*}\times\frac{\int \tfrac{dq}{q_*}\tfrac{q}{q_*}\left( P_{+}^2(q)- P_{-}^2(q)\right)}{\int \tfrac{dq}{q_*}\tfrac{q^2}{q^2_*}\left( P_{+}^2(q)+ P_{-}^2(q)+12P_{+}(q)P_{-}(q)\right)}\,,
\end{align}
where we neglected potential ${\cal O}(1)$ factors arising from the time integral in Eq.~\eqref{eq:omegaGW}. We also find that a similar argument applies to the scalar-tensor interactions.
This shows that there is a universal IR scaling for the chirality fraction for parity-violating causal GW sources, which is always linearly proportional to $k$. We now look at two particular examples of a peaked power spectrum as a consistency check of our calculations.

\subsection{Particular example: peaked spectrum}

To deal with concrete examples of the power spectrum, we find it more convenient to work with dimensionless quantities. We introduce the dimensionless power spectrum ${\cal P}_s(k)$, that is
\begin{align}\label{eq:dimensionfullP}
P_{s}(k)=\frac{2\pi^2}{k^3}{\cal P}_s(k)\,,
\end{align}
and the dimensionless internal momenta, as in Ref.~\cite{Kohri:2018awv}, given by
\begin{align}
v=q/k \quad {\rm and}\quad u = q /|\bk-\bq|\,.
\end{align}
With the above redefinitions, the GW spectrum \eqref{eq:omegaGW} for a Gaussian source as in Eq.~\eqref{eq:sourcegaussian2} reads
\begin{align}\label{eq:omegaGW22}
        \Omega^\lambda_\mathrm{GW}(k) = \, \frac{1}{6 a^2{\cal H}^2 \mpl^4}\int_{0}^{\eta} d\eta_1\int_{0}^{\eta}& d\eta_2 \,
        a^3 (\eta_1)a^3 (\eta_2)\cos \left(k(\eta_1-\eta_2)\right)\nonumber\\&\times \sum_{s_1,s_2}\int_0^\infty \frac{dv}{v^2}\int_{|1-v|}^{1+v}\frac{du}{u^2} \big|Q^\lambda_{s_1s_2}(v,u)\big|^2{\cal P}_{s_1}(vk){\cal P}_{s_2}(uk)\,,
\end{align}
where
\begin{align}
\big|Q^\lambda_{s_1s_2}(v,u)\big|=\frac{\left(2 v+\lambda s_1 \left(1-u^2+v^2\right)\right)^2 \left(2 u+\lambda s_2
   \left(1+u^2-v^2\right)\right)^2}{16 u v}\,.
\end{align}

We now proceed to study a Dirac delta and a log-normal power spectrum. Note that since we are mostly interested in the IR scaling laws and the chirality fraction in the IR regime, we will drop most of the irrelevant factors in Eq.~\eqref{eq:omegaGW22} and focus entirely on the overall $k$ dependence.

\subsubsection{Dirac delta spectrum}

Let us assume that the power spectrum of each polarization only has power at a single characteristic wavenumber $k^*$, namely
\begin{equation}\label{eq:DiracP}
    \mathcal{P}_\pm(xk) = \mathcal{A}_\pm\delta\left(\ln\frac{xk}{k^*}\right)=\mathcal{A}_\pm\frac{1}{\kappa}\delta\left(x-1/\kappa\right) \quad{\rm with} \quad\kappa=k/k^*\,\,.
\end{equation}
Plugging this back into the GW energy density parameter \eqref{eq:omegaGW22}, we find that
\begin{align}
    \Omega_{\text{GW}}^\pm & \propto {\kappa^2} \left(\big|Q^\pm_{++}(\kappa)\big|^2\mathcal{A}_+^2+\big|Q^\pm_{--}(\kappa)\big|^2\mathcal{A}_-^2+2\big|Q^\pm_{+-}(\kappa)\big|^2\mathcal{A}_+\mathcal{A}_-\right)\,, \label{eq:Omega_GW_Dirac}
\end{align}
where
\begin{align}\label{eq:QsDirac}
\big|Q^\pm_{++}(\kappa)\big|^2=\left(\frac{\kappa\pm 2}{4}\right)^4\quad,\quad\big|Q^\pm_{--}(\kappa)\big|^2=\left(\frac{\kappa\mp 2}{4}\right)^4\quad{\rm and}\quad \big|Q^\pm_{+-}(\kappa)\big|^2=\left(\frac{\kappa^2-4}{16}\right)^2\,.
\end{align}
The chirality fraction is then given by
\begin{align} \label{eq:Pi_GW_Dirac}
    \Pi_{\text{GW}} &=\frac{\left(\big|Q^+_{++}(\kappa)\big|^2-\big|Q^+_{--}(\kappa)\big|^2\right)\left({\mathcal{A}_+}^2-{\mathcal{A}_-}^2\right)}{\left(\big|Q^+_{++}(\kappa)\big|^2+\big|Q^+_{--}(\kappa)\big|^2\right)\left({\mathcal{A}_+}^2+{\mathcal{A}_-}^2\right)+2\big|Q^+_{+-}(\kappa)\big|^2\mathcal{A}_+\mathcal{A}_-} \,.
\end{align}
In the IR limit, that is $\kappa\ll1$, we find that
\begin{align}\label{eq:chiralitydeltadirac}
\Pi_{\rm GW}\approx 2\kappa \times \frac{{\mathcal{A}_+}^2-{\mathcal{A}_-}^2}{({\mathcal{A}_+}+{\mathcal{A}_-})^2}
\end{align}

As one can see from Eq.~\eqref{eq:Omega_GW_Dirac}, the energy density parameter in the IR limit scales as $\Omega^{\pm}_{\rm GW}\propto \kappa^2$, in accordance with \cite{Ananda:2006af,Saito:2008jc} and noted in Ref.~\cite{Cai:2019cdl} as an exception to the $k^3$ causal IR tail. Nevertheless, the chirality fraction \eqref{eq:chiralitydeltadirac} scales linearly with $\kappa$. We also see that the maximal chirality fraction in the IR tail is $\Pi_{\rm GW}\approx \pm 2\kappa$, respectively for  ${\cal A}_-=0$ and ${\cal A}_+=0$.

\subsubsection{Log-normal spectrum}

Let us now consider a finite width of the power spectrum with a log-normal given by
\begin{equation}\label{eq:lognormalP}
    \mathcal{P}_\pm(xk)=\frac{\mathcal{A}_{\pm}}{\sqrt{2\pi}\mu}e^{-\frac{1}{2\mu^2}\ln^2(xk)} \,,
\end{equation}
where $\mu$ is the logarithmic width of the spectrum. As shown in Ref.~\cite{Pi:2020otn}, the $\kappa$ terms factor out of the log-normal functions under the following change of variables,
\begin{align}
    s=\frac{1}{\sqrt{2}}\ln(uv) \quad \text{and} \quad t=\frac{1}{\sqrt{2}}\ln(\frac{u}{v}) \,.
\end{align}
The energy density parameter then reads
\begin{align}
        \Omega^\pm_{\text{GW}} \propto & \, 2\pi\kappa^2e^{\mu^2}\sum_{s_1,s_2}\frac{\mathcal{A}_{s_1}}{\sqrt{2\pi}\mu}\frac{\mathcal{A}_{s_2}}{\sqrt{2\pi}\mu}\int_{-\infty}^{+\infty}ds\,
        e^{-\frac{1}{2\mu^2}\left(s+\sqrt{2}\left(\ln\kappa+\mu^2\right)\right)^2}\int_{\chi(s)}^{\xi(s)}dt \,
        e^{-\frac{t^2}{2\mu^2}}\big|Q^\pm_{s_1s_2}(s,t)\big|^2 \,,
\end{align}
where the integration boundaries $\chi$ and $\xi$ are defined by
\begin{align}
    \chi(x)=\sqrt{2}\Re\left(\arccosh\left(\frac{1}{2}e^{-\frac{x}{\sqrt{2}}}\right)\right) \qquad \text{and} \qquad
    \xi(x)=\sqrt{2}\arcsinh\left(\frac{1}{2}e^{-\frac{x}{\sqrt{2}}}\right) \,,
\end{align}
respectively.

In the IR limit, we have that $t\to0$ and the integral over $t$ may be crudely approximated by taking $\big|Q^\pm_{s_1s_2}(s,t=0)\big|$ and performing the remaining Gaussian integral in $t$ to yield an Erf-function. By further setting $x=s+\sqrt{2}\left(\ln\kappa+\mu^2\right)$, this leads to
\begin{align}
    \Omega^\pm_{\text{GW}} \propto&\,\frac{\kappa^2e^{\mu^2}\pi}{\mu^2}\sum_{s_1,s_2}{\mathcal{A}_{s_1}}{\mathcal{A}_{s_2}} \int_{-\infty}^{+\infty}dx\,e^{-\frac{1}{2\mu^2}x^2}\Big|Q^\pm_{s_1s_2}\left(x-\sqrt{2}\left(\ln\kappa+\mu^2\right),0\right)\Big|^2 \nonumber\\
    & \phantom{\times\int_{-\infty}^{+\infty}dx}\times\sqrt{\frac{\pi}{2}}\mu\left[\Erf\left(\tfrac{\xi\left(x-\sqrt{2}\left(\ln\kappa+\mu^2\right)\right)}{\sqrt{2}\mu}\right)-\Erf\left(\tfrac{\chi\left(x-\sqrt{2}\left(\ln\kappa+\mu^2\right)\right)}{\sqrt{2}\mu}\right)\right] \\
    \propto&\, \frac{\kappa^2e^{\mu^2}\pi}{\mu^2}\sum_{s_1,s_2}\mathcal{A}_{s_1}\mathcal{A}_{s_2}\int_{-\infty}^{+\infty}dx\,e^{-\frac{1}{2\mu^2}x^2}\mathcal{K}_{s_1s_2}^\pm(x) \,,
\end{align}
where we defined the integration kernel by
\begin{align}\label{eq:bigK}
    \mathcal{K}_{s_1s_2}^\pm(x)=&\,\frac{1}{256\sqrt{2}}
    \kappa e^{5\left(\mu^2 - \frac{x}{\sqrt{2}}\right)} \times
    \begin{cases}
        \left(\kappa\mp{2}e^{\frac{x}{\sqrt{2}}}e^{-\mu^2}\right)^4 \quad &(s_1,s_2=+,+)\\
        \left(\kappa\pm 2e^{\frac{x}{\sqrt{2}}}e^{-\mu^2}\right)^4 \quad &(s_1,s_2=-,-)\\
        \left(\kappa^2-{4}e^{\sqrt{2}x}e^{-2\mu^2}\right)^2 \quad &(s_1,s_2=\pm,\mp)
    \end{cases} \,.
\end{align}
As one can see from Eq.~\eqref{eq:bigK}, the only parity-violating contributions in the integration kernel, i.e.~those that differ for ``$+$" and ``$-$" polarizations, are the first two. Thus, one may neglect the others when computing the parity-violation parameter. In the IR limit, the parity-violating part and total GW spectrum respectively scale as
\begin{align}
    \Omega^{+}_\text{GW}-\Omega^{-}_\text{GW} &\propto -\frac{\kappa^4e^{3\mu^2}\pi}{2\sqrt{2}\mu^2}\int_{-\infty}^{+\infty}dx \, e^{-\sqrt{2}x-\frac{1}{2\mu^2}x^2}\left({\mathcal{A}^+}^2-{\mathcal{A}^-}^2\right) \,,\\
    \Omega^{+}_\text{GW}+\Omega^{-}_\text{GW}&\propto \frac{\kappa^3e^{2\mu^2}\pi}{4\sqrt{2}\mu^2}\int_{-\infty}^{+\infty}dx \,e^{-\frac{x}{\sqrt{2}}-\frac{1}{2\mu^2}x^2}\left({\mathcal{A}^+}^2+{\mathcal{A}^-}^2+4\mathcal{A}^+\mathcal{A}^-\right) \,, \label{eq:Omega_GW_lognormal}
\end{align}
thus again yielding a chirality fraction linear in $\kappa$,
\begin{equation}
    \Pi_\text{GW}\propto\kappa \,.
\end{equation}
Note that, in the log-normal case, the total energy density parameter Eq.~\eqref{eq:Omega_GW_lognormal} scales as $\kappa^3$, as found in \cite{Cai:2019cdl}, but the chirality fraction still scales linearly in $\kappa$, as in the Dirac delta case.

\section{Parity violation in scalar-induced GWs}\label{sec:Trispectrum}

After having studied parity violation in GW backgrounds sourced by Gaussian vector fields in Sec.~\ref{sec:ir-vector}, we now turn to study scalar-induced GWs (SIGWs). SIGWs arise at second order in cosmological perturbation theory, where the scalar and tensor sectors of space-time perturbations mix \cite{Tomita:1967wkp,Matarrese:1993zf,Matarrese:1997ay} and scalar perturbations source gravitational waves \cite{Ananda:2006af,Baumann:2007zm}. SIGW have received significant attention in the recent literature due to their close connection to primordial black hole (PBH) formation \cite{Saito:2008jc, Yuan:2021qgz}, as well as their potential to probe the primordial power spectrum on small scales. For a review, see Ref.~\cite{Domenech:2021ztg}. Note that when dealing with scalar-induced GWs, we can take into account all transfer functions of linear fluctuations that we have neglected in the previous section. Nevertheless, since most of the generation occurs at the horizon crossing of the scalar field fluctuations, it is effectively a finite-time source.

Scalar perturbations arising from quantum fluctuations during inflation are typically close to being Gaussian; however, small departures from Gaussianity -- generically referred to as non-Gaussianity -- may arise due to gravitational interactions \cite{Maldacena:2002vr,Koyama:2010xj}.
The presence of such non-Gaussianities may have important observational consequences \cite{Domenech:2021ztg,Atal:2021jyo,Yuan:2021qgz,Adshead:2021hnm,Ragavendra:2021qdu,Perna:2024ehx,Iovino:2024sgs,LISACosmologyWorkingGroup:2025vdz,Perna:2026szd}, in particular for the formation of PBHs and the interpretation of the associated SIGW spectra \cite{Iovino:2025cdy}.

\subsection{Parity violation in SIGW}\label{subsec:intro-PV-SIGW}

Although SIGWs are sourced by a scalar field and, therefore, the two-point correlator has no parity, it is possible that the four-point correlator, the trispectrum, contains parity-violating terms \cite{Cabass:2022rhr,Garcia-Saenz:2023zue}. Since we cannot draw general conclusions without specifying the shape of the trispectrum, we use the templates proposed by Refs.~\cite{Ragavendra:2025svk,Caporali:2026qhc}, explicitly shown in Eqs.~\eqref{eq:template1} and \eqref{eq:Todd} below. We believe these are well-motivated and suited templates to study the IR tail of the SIGW spectrum, since they are based on the collapsed shape of the trispectrum of curvature fluctuations \cite{Byrnes:2006vq}. As discussed below Eq.~\eqref{eq:EforQs}, it is precisely the collapsed limit of the trispectrum which contributes to the IR tail of SIGWs at leading order. Although the SIGWs from these templates were studied in Refs.~\cite{Ragavendra:2025svk,Caporali:2026qhc}, we extend their analysis by studying the chirality fraction in the IR limit and show that they also exhibit the universal linear scaling as the Gaussian case discussed in Sec.~\ref{sec:ir-gaussian}.

The connected trispectrum $\mathcal{T}(\bk_1,\bk_2,\bk_3,\bk_4)$ of the comoving curvature perturbation is defined by \cite{Ragavendra:2025svk,Caporali:2026qhc}
\begin{equation}\label{eq:trispectrum_def}
    \langle\mathcal{R}_{\bk_1}\mathcal{R}_{\bk_2}\mathcal{R}_{\bk_3}\mathcal{R}_{\bk_4}\rangle=\mathcal{T}(\bk_1,\bk_2,\bk_3,\bk_4)\delta^{(3)}(\bk_1+\bk_2+\bk_3+\bk_4) \,.
\end{equation}
Using the same definition for $\mathcal{T}(\bk_1,\bk_2,\bk_3,\bk_4)$ as in Refs.~\cite{Ragavendra:2025svk,Caporali:2026qhc} but in the convention used in our paper, we have that
\begin{equation}\label{eq:trispectrum_def2}
    \langle\mathcal{R}_{\bk_1}\mathcal{R}_{\bk_2}\mathcal{R}^*_{\bk_3}\mathcal{R}^*_{\bk_4}\rangle=\mathcal{T}(\bk_1,\bk_2,-\bk_3,-\bk_4)\delta^{(3)}(\bk_1+\bk_2-\bk_3-\bk_4) \,.
\end{equation}
For purely Gaussian curvature perturbations, Eq.~\eqref{eq:trispectrum_def} reduces to a sum of products of the power spectrum (see Eq.~\eqref{eq:four-point-a}), while in the general non-Gaussian case, the functional form of $\mathcal{T}$ is a priori unknown.

Adding the Gaussian and connected trispectrum contributions, the SIGW spectral energy density can be expressed as \cite{Adshead:2021hnm,Perna:2024ehx}
\begin{align}\label{eq:omegaSIGWs}
        \Omega^{\pm}_\text{GW} = \frac{1}{3}&
        \int_0^\infty dv_q\int_{|1-v_q|}^{1+v_q}
        \frac{du_q}{(u_qv_q)^2} \, \big|Q^\pm(k,v_q)\big|^2\tilde{I}_1(v_q,u_q)\mathcal{P}_{\cal R}(kv_q)\mathcal{P}_{\cal R}(ku_q) \nonumber\\
        & +\frac{k^9}{24\pi^4}
        \iint_0^\infty dv_qdv_l\int_{|1-v_q|}^{1+v_q}du_q\int_ {|1-v_l|}^{1+v_l}d u_l\,v_qv_lu_qu_l \int_0^{2\pi}d\psi\nonumber \\
        & \qquad \qquad \times \Big\{Q^\pm(k,v_q){Q^\pm}^*(k,v_l)\tilde{I}_2(v_q,v_l,u_q,u_l)\mathcal{T}(\bq,\bk-\bq,-\bl,-\bk+\bl)\Big\} \,,
\end{align} 
where $\psi=\varphi_q-\varphi_l$ and we defined
\begin{align}
v_q=q/k\quad,\quad v_l=l/k\quad,\quad u_q=|\bk-\bq|/k\quad{\rm and}\quad u_l=|\bq-\bl|/k\,.
\end{align}
Note that, since we assumed a radiation-dominated background, there is no time dependence in the energy density fraction in the sub-Hubble limit, that is, $k\eta\gg1$.
The first term, involving the power spectrum $\mathcal{P}_{\cal R}$, is the usual term arising from purely Gaussian fluctuations, and we will refer to this term as the ``Gaussian part", while the second term, involving the trispectrum $\mathcal{T}$, arises only in the presence of non-Gaussianity and may carry parity violation.

The geometrical factors $Q^{\pm}$ in Eq.~\eqref{eq:omegaSIGWs} encode the projection of the polarization tensors on the tensor modes' momenta and are defined by
\begin{align}
    Q^\pm(k,v_{q})=e^\pm_{ij}(\bk)\frac{q^iq^j}{k^2} \,,
\end{align}
and similarly for $Q^\pm(k,v_l)$.
The integration kernels in Eq.~\eqref{eq:omegaSIGWs} are given by \cite{Adshead:2021hnm}
\begin{align}
        \tilde{I}_1(u,v)&=I_A^2(u,v) \,\left[I_B^2(u,v)+\pi^2I_C^2(u,v)\right] \,,
        \\
        \tilde{I}_2(u_q,v_q,u_l,v_l)&=I_A(u_q,v_q)I_A(u_l,v_l)\, \left[I_B(u_q,v_q)I_B(u_l,v_l)+\pi^2I_C(u_q,v_q)I_C(u_l,v_l)\right] \,,
\end{align}
where
\begin{align}
        I_A(u,v)&=\frac{3}{4}\frac{u^2+v^2-3}{u^3v^3} \,,
        \\
        I_B(u,v)&=-4uv+(u^2+v^2-3)\ln\left|\frac{3-(u+v)^2}{3-(u-v)^2}\right| \,,
        \\
        I_C(u,v)&=(u^2+v^2-3)\Theta(u+v-3) \,.
\end{align}

The trispectrum in Eq.~\eqref{eq:omegaSIGWs}, $\mathcal{T}(\bq,\bk-\bq,-\bl,-\bk+\bl)$, is in general a function of the azimuthal angle difference $\psi=\varphi_q-\varphi_l$, by symmetry arguments \cite{Garcia-Saenz:2023zue}. If it's an even function of $\psi$, it contributes to the total GW energy density but has vanishing chirality. We call this component $\mathcal{T}_{\rm even}(\bq,\bk-\bq,-\bl,-\bk+\bl)$. If it's an odd function of $\psi$, it only contributes to the helicity power difference and yields non-zero chirality. We call this component $\mathcal{T}_{\rm odd}(\bq,\bk-\bq,-\bl,-\bk+\bl)$. 

Unfortunately, we cannot draw conclusions about the $ k$-dependence based on symmetries alone. For this reason, we use the two templates proposed in Ref.~\cite{Ragavendra:2025svk}. For simplicity, we focus on the first template in the main text. We discuss the second template in App.~\ref{app:template_2}, since it yields similar results. Explicitly, the even and odd components of the first template are respectively given by
\begin{align}\label{eq:template1}
    \mathcal{T}_{\text{even}}(\bk_1,\bk_2,\bk_3,\bk_4) & = 2\tau_{\text{NL}}P_{\cal R}(k_1)P_{\cal R}(k_3)P_{\cal R}(|\bk_1+\bk_2|)+\text{11 permutations}\,,\\
    \mathcal{T}_{\text{odd}}(\bk_1,\bk_2,\bk_3,\bk_4) & = i\tilde{\tau}_{\text{NL}}\beta(\widehat{\bk_1+\bk_2},\hat{\bk}_1,\hat{\bk}_3)P_{\cal R}(k_1)P_{\cal R}(k_3)P_{\cal R}(|\bk_1+\bk_2|)+\text{23 permutations}\,,\label{eq:Todd}
\end{align}
where $P_{\cal R}(k)$ is the dimensionful power spectrum (see, e.g., Eq.~\eqref{eq:dimensionfullP}), and we introduced
\begin{align}\label{eq:beta}
\beta\left(\hat{\bk}_1,\hat{\bk}_2,\hat{\bk}_3\right)=\hat{\bk}_1\cdot\left(\hat{\bk}_2\times\hat{\bk}_3\right) \,.
\end{align}
Note that we use $\tau_{\rm NL}$ instead of $g_{\rm NL}$ as Ref.~\cite{Ragavendra:2025svk} to denote the coefficient of the trispectrum to match standard literature on trispectrum shapes. $\tau_{\rm NL}$ is often used as the coefficient of the collapsed shape of the trispectrum \cite{Seery:2006vu,Byrnes:2006vq}, that is, $k_1+k_2=-(k_3+k_4)\to 0$ in Eq.~\eqref{eq:trispectrum_def}), while $g_{\rm NL}$ is used for the cubic local expansion of non-Gaussianity \cite{Byrnes:2006vq} (often called the squeezed shape); see \cite{Perna:2024ehx} for the SIGWs from $g_{\rm NL}$. 

The total trispectrum is given by $\mathcal{T}=\mathcal{T}_{\text{odd}}+\mathcal{T}_{\text{even}}$. Note that, while the templates for the even and odd components have the same dependence in the power spectrum, the function $\beta$ \eqref{eq:beta} switches sign under $\bk_i\to-\bk_i$ and thus carries the parity violation of $\mathcal{T}_{\text{odd}}$.   Note that due to the symmetries of the integration plane in Eq.~\eqref{eq:omegaSIGWs}, the total number of permutations reduces to $12$ for the odd trispectrum.

Since the collapsed shape \eqref{eq:template1} is equivalent to the contribution of quadratic local non-Gaussinity (often denoted by $f_{\rm NL}$) to the four-point correlator \cite{LISACosmologyWorkingGroup:2025vdz}, we can map the contributions to the GW energy density from $\tau_{\rm NL}$ to the ones of Refs.~\cite{Adshead:2021hnm,Perna:2024ehx} at order $f^2_{\rm NL}$.\footnote{For fluctuations generated during inflation one has the Suyama-Yamaguchi consistency relation which states that $\tau_{\rm NL}\geq (6f_{\rm NL}/5)^2$ \cite{Suyama:2007bg,Suyama:2010uj}.} We review this equivalence in App.~\ref{app:F_NL-expansion}. The independent contributions are denoted as $C$ and $Z$ terms in Refs.~\cite{Adshead:2021hnm,Perna:2024ehx}. For convenience, we will also use the notation $C$- and $Z$-like terms for the odd trispectrum as the scale dependence is the same as the corresponding terms in the even trispectrum, apart from the $\beta$ prefactor \eqref{eq:beta}.

Once we specified the shape of the trispectrum, we could take the IR limit for a general shape of the power spectrum, since we have all the explicit momentum dependence in Eq.~\eqref{eq:omegaSIGWs}. However, since the main purpose of the paper is to study the scaling laws, we simplify the calculations by restricting to two particular, analytically viable, examples considered in the previous section: the Dirac delta \eqref{eq:DiracP} and the log-normal \eqref{eq:lognormalP} spectra. From the latter, similar conclusions can be extended to any shape of the primordial spectrum as long as it is peaked or has an IR cut-off. In the next subsections, we focus only on IR scaling and therefore drop the numerical coefficients for simplicity.

\subsection{Dirac delta spectrum\label{sec:diracdeltaSIGWs}}

For the Dirac delta spectrum given by Eq.~\eqref{eq:DiracP} but with amplitude ${\cal A}_R$, out of the 12 permutations for the even trispectrum, only 8 are non-vanishing after the integration over the difference of the two azimuthal angles $\psi$. Below, we present the main results and refer the reader to App.~\ref{app:dirac-delta-energy-density} for details.

On the one hand, for the Gaussian term, we find that
\begin{equation}\label{eq:diracgausigws}
    \Omega_{\text{GW}}^{\text{gaus}} \propto \mathcal{A}^2_\mathcal{R}\kappa^2\ln^2\kappa \,,
\end{equation}
in agreement with Refs.~\cite{Cai:2019cdl,Yuan:2019wwo}. Note that the logarithmic running in Eq.~\eqref{eq:diracgausigws} stems from a secular sub-horizon growth of the GW amplitude but disappears once dissipative effects are taken into account \cite{Domenech:2025bvr}. 

On the other hand, we find that the $C$ terms scale as $\kappa^2$ (even) and $\kappa^3$ (odd), due to a $1/\kappa^2$ factor coming from the boundary of the integral. The resulting GW energy spectra due to the $C$ terms schematically take the form of
\begin{align}\label{eq:dirac-energy-density}
        \Omega_{\text{GW}}^{\text{even}}[C] & \propto\mathcal{A}^3_\mathcal{R} \tau_{\text{NL}}\kappa^2\ln^2\kappa \,,\\
        \Omega_{\text{GW}}^{\text{odd}}[C] & \propto\mathcal{A}^3_\mathcal{R}\tilde{\tau}_{\text{NL}}\kappa^3\ln^2\kappa\,,
\end{align}
for the even and odd parts, respectively. The $Z$ terms are subdominant with respect to the $C$ terms and are roughly given by 
\begin{align}\label{eq:evenZdirac}
        \Omega_{\text{GW}}^{\text{even}}[Z]  &\propto - \mathcal{A}^3_\mathcal{R}\tau_{\text{NL}}\kappa^3\ln^2\kappa \,, \\
        \Omega_{\text{GW}}^{\text{odd}}[Z] & \propto \mathcal{A}^3_\mathcal{R}\tilde{\tau}_{\text{NL}}\kappa^4\ln^2\kappa \,. \label{eq:oddZdirac}
\end{align}
Although the $Z$ contribution to the total GW energy density \eqref{eq:evenZdirac} is negative, it is subdominant, so that the total energy density stays positive definite. These scalings are consistent with the ones found by Ref.~\cite{Adshead:2021hnm}.

From Eqs.~\eqref{eq:diracgausigws} to \eqref{eq:oddZdirac}, we conclude that the chirality fraction in the IR scales as
\begin{equation}
    \Pi_{\text{GW}}(\kappa\ll 1) = \frac{\Omega_{\text{GW}}^{\text{odd}}}{\Omega_{\text{GW}}^{\text{gaus}}+\Omega_{\text{GW}}^{\text{even}}} \propto \kappa\,.
\end{equation}
The linear dependence arises from the ratio of the odd and even (or Gaussian) parts. Note that, if generated during standard inflation, one generally expects the Gaussian component to dominate over the connected trispectrum \cite{Garcia-Saenz:2022tzu}. That would imply a small amount of parity-violation, and a small chirality fraction, namely $\Pi_{\text{GW}}\approx \Omega_{\text{GW}}^{\text{odd}}/\Omega_{\text{GW}}^{\text{gaus}}\ll1$. Nevertheless, the universal linear scaling of the chirality fraction derived here applies regardless of any hierarchy between components.

\subsection{Log-normal spectrum}

We now proceed to study the case of the log-normal spectrum given by Eq.~\eqref{eq:lognormalP} but with amplitude ${\cal A}_{\cal R}$. The calculations of the trispectrum-induced GWs from Eq.~\eqref{eq:template1} are fairly involved but involve similar steps as in the Dirac delta case for both $\bl$ and $\bq$, detailed in App.~\ref{app:dirac-delta-energy-density}. As the calculations are not particularly illuminating and we are interested in the IR scaling, we provide a summary of the main steps below. We also focus mainly on the even component because it is shorter. The odd component follows the same calculations but multiplied by the factor $\beta$ \eqref{eq:beta}. In the IR, the factor $\beta$ may bring in an additional linear dependence in $\kappa$ with respect to the even component.

Let us denote with $i=[1,12]$ the number of permutations in the templates \eqref{eq:template1} and \eqref{eq:Todd}. For clarity, we explicitly write down the order of the permutations taken in Eq.~\eqref{eq:permutations_set}. We also label with $(i)$ the corresponding component to the trispectrum induced GWs, namely
\begin{align}\label{eq:ln-small-omega}
        \Omega^{\text{even},(i)}_\text{GW} &= 16\pi\tau_{\text{NL}}\iint_0^\infty\frac{d\alpha}{\alpha}\frac{d\beta}{\beta}\int_0^{2\pi}d\psi~\omega_\text{GW}^{\text{even},(i)} \,,\\
        \Omega^{\text{odd},(i)}_\text{GW} &= 16\pi i\tilde{\tau}_{\text{NL}}\iint_0^\infty\frac{d\alpha}{\alpha}\frac{d\beta}{\beta}\int_0^{2\pi}d\psi~\omega_\text{GW}^{\text{odd},(i)} \,,
\end{align}
where we defined
\begin{align}
    \alpha=e^{x_l/\sqrt{2}} \quad \text{and} \quad \beta=e^{x_q/\sqrt{2}} \,.
\end{align}
We give the explicit expressions of the relevant $\omega_\text{GW}^{\text{even},(i)}$ and $\omega_\text{GW}^{\text{odd},(i)}$ shortly below.

First, we find that contributions from the permutations $\{1,2,11,12\}$ are highly suppressed in the IR limit, as they are all proportional to $\exp \left(-\ln^2\kappa/\left(2\mu^2\right)\right)/\mu$.
The result for the log-normal case is thus consistent with the Dirac delta limit, that is, when $\mu\to 0$, for which those terms vanished. The remaining permutations can all be related to a single independent contribution. For instance, an exchange of the integration variables $\alpha$ and $\beta$ leads to the identification
\begin{align}
    \begin{split}
        & \omega_\text{GW}^{(3)}=\omega_\text{GW}^{(8)}\quad;\quad\omega_\text{GW}^{(4)}=\omega_\text{GW}^{(7)}\\
        & \omega_\text{GW}^{(5)}=\omega_\text{GW}^{(10)}\quad;\quad\omega_\text{GW}^{(6)}=\omega_\text{GW}^{(9)}\\
    \end{split}
\end{align}
for both even and odd parts. Furthermore, we find that the permutations 3 to 6 are linked by
\begin{equation}
    \setlength{\arraycolsep}{17pt}
    \begin{array}{ll}
        \omega_\text{GW}^{\text{even},(3)} = \left(\frac{\alpha}{\beta}\right)^3 e^{\frac{\ln^2\alpha-\ln^2\beta}{2\mu^2}}\omega_\text{GW}^{\text{even}(4)}\,, &
       \omega_\text{GW}^{\text{odd},(3)} = -\left(\frac{\alpha}{\beta}\right)^4 e^{\frac{\ln^2\alpha-\ln^2\beta}{2\mu^2}}\omega_\text{GW}^{\text{odd}(4)}\,, \\
       \omega_\text{GW}^{\text{even},(5)} = \left(\frac{\alpha}{\beta}\right)^3 e^{\frac{\ln^2\alpha-\ln^2\beta}{2\mu^2}}\omega_\text{GW}^{\text{even}(6)}\,, &
       \omega_\text{GW}^{\text{odd},(5)} = -\left(\frac{\alpha}{\beta}\right)^4 e^{\frac{\ln^2\alpha-\ln^2\beta}{2\mu^2}}\omega_\text{GW}^{\text{odd}(6)}\,,
    \end{array}
\end{equation}
and, lastly,
\begin{align}
        \omega_\text{GW}^{\text{even},(3)} &= \left(\tfrac{g(\psi)}{f(\psi)}\right)^{3/2}e^{\frac{\ln^2 \sqrt{g(\psi)}-\ln^2 \sqrt{f(\psi)}}{2\mu^2}}\omega_\text{GW}^{\text{even}(5)}\,, \\
       \omega_\text{GW}^{\text{odd},(3)} &= -\left(\tfrac{g(\psi)}{f(\psi)}\right)^{3/2}e^{\frac{\ln^2 \sqrt{g(\psi)}-\ln^2 \sqrt{f(\psi)}}{2\mu^2}}\omega_\text{GW}^{\text{odd}(5)} \,.
\end{align}

In the end, we obtain that both the even and odd energy densities can be expressed by the sole contributions $\omega_\text{GW}^{\text{even},(3)}$ and $\omega_\text{GW}^{\text{odd},(3)}$, which are respectively given by
\begin{align} \label{eq:ln-energy-density}
        \Omega_\text{GW}^\text{even} &= 16\pi\tau_{\text{NL}}\iint_0^\infty\frac{d\alpha}{\alpha}\frac{d\beta}{\beta}\int_0^{2\pi}d\psi~\omega_\text{GW}^{\text{even},(3)}\left[1+\left(\frac{\beta}{\alpha}\right)^3e^{\frac{\ln^2\beta-\ln^2\alpha}{2\mu^2}}\right]\nonumber\\&\qquad\qquad\qquad\qquad\qquad\qquad\qquad\qquad\qquad\times\left[1+\left(\frac{f(\psi)}{g(\psi)}\right)^{3/2}e^{\frac{\ln^2 \sqrt{f(\psi)}-\ln^2 \sqrt{g(\psi)}}{2\mu^2}}\right] \,,
\end{align}
and
\begin{align}
        \Omega_\text{GW}^\text{odd} &= 16\pi i\tilde{\tau}_{\text{NL}}\iint_0^\infty\frac{d\alpha}{\alpha}\frac{d\beta}{\beta}\int_0^{2\pi}d\psi~\omega_\text{GW}^{\text{odd},(3)}\left[1-\left(\frac{\beta}{\alpha}\right)^4e^{\frac{\ln^2\beta-\ln^2\alpha}{2\mu^2}}\right]\nonumber\\&\qquad\qquad\qquad\qquad\qquad\qquad\qquad\qquad\qquad\times\left[1-\left(\frac{f(\psi)}{g(\psi)}\right)^{3/2}e^{\frac{\ln^2 \sqrt{f(\psi)}-\ln^2 \sqrt{g(\psi)}}{2\mu^2}}\right] \,,
\end{align}
where we defined
\begin{align} \label{eq:function-f-def}
    f(\psi)&=\alpha^2+\beta^2+\frac{\kappa^2}{2}-2\alpha\beta\sqrt{1-\frac{\kappa^2}{4\alpha^2}}\sqrt{1-\frac{\kappa^2}{4\beta^2}}\cos\psi \,, \\
     \label{eq:function-g-def}
    g(\psi)&=\alpha^2+\beta^2+\frac{\kappa^2}{2}+2\alpha\beta\sqrt{1-\frac{\kappa^2}{4\alpha^2}}\sqrt{1-\frac{\kappa^2}{4\beta^2}}\cos\psi \,,
\end{align}
and
\begin{align}\label{eq:ln-parameters_1}
        & \omega_\text{GW}^{\text{even},(3)}=36\sqrt{2}\pi^{5/2}\mathcal{A}^3_\mathcal{R}\kappa^3\ln\frac{\alpha}{\kappa}\ln\frac{\beta}{\kappa}\left(\frac{\alpha}{\beta}\right)^3\frac{1}{\mu^3}e^{-\frac{\ln^2\beta}{\mu^2}}e^{-\frac{\ln^2\sqrt{f(\psi)}}{2\mu^2}}\frac{\cos(2\psi)}{f^{3/2}(\psi)} \,,\\
        & \omega_\text{GW}^{\text{odd},(3)}=18i\pi^{5/2}\mathcal{A}^3_\mathcal{R}\kappa^4\ln\frac{\alpha}{\kappa}\ln\frac{\beta}{\kappa}\sqrt{4-\left(\frac{\kappa}{\alpha}\right)^2}\sqrt{4-\left(\frac{\kappa}{\beta}\right)^2}\frac{\alpha^3}{\beta^4}\frac{1}{\mu^3}e^{-\frac{\ln^2\beta}{\mu^2}}e^{-\frac{\ln^2\sqrt{f(\psi)}}{2\mu^2}}\frac{\cos\psi\sin^2\psi}{f^{3/2}(\psi)} \,.
        \label{eq:ln-parameters_2}
\end{align}
We note that the $C$ and $Z$ terms can be recovered by taking the first bracket of both of these expressions. Concretely, multiplying by the factor $1$ gives the $Z$ terms, whereas the factors arising from $\left(\frac{\beta}{\alpha}\right)^4e^{\frac{\ln^2\beta-\ln^2\alpha}{2\mu^2}}$ are the $C$ terms. We checked that the log-normal distribution recovers the Dirac delta one in the asymptotic limit of vanishing width, namely $\mu \rightarrow 0$.\footnote{The $\mu\to 0$ limit of the expressions Eqs.~\eqref{eq:ln-parameters_1} and \eqref{eq:ln-parameters_2}, sets $f(\psi)$ and $g(\psi)$ to 1. The same limit also sets $\alpha$ or $\beta$ or $\alpha$ and $\beta$ to $1$, depending on the permutation. By doing so, we recovered the $C$ and $Z$ terms derived in Sec.~\ref{subsec:intro-PV-SIGW}.}

Now, we investigate the IR limit of the SIGW spectrum. From Eqs.~\eqref{eq:ln-parameters_1} and \eqref{eq:ln-parameters_2} we see that in the $\kappa\to 0$ limit the dominant components respectively are $\Omega_\text{GW}^\text{even}\propto \kappa^3\ln^2\kappa$ and $\Omega_\text{GW}^\text{odd}\propto \kappa^4\ln^2\kappa$. The remaining terms inside the integral are $\kappa$-independent and, therefore, just give a number after the definite integrals. Furthermore, we checked that the Gaussian component in Eq.~\eqref{eq:omegaSIGWs} for the log-normal spectrum scales as
\begin{equation}
    \Omega_\text{GW}^\text{gaus}\propto \mathcal{A}_\mathcal{R}^2\kappa^3\ln^2\kappa\frac{1}{\mu^2}\int_0^\infty \frac{d\beta}{\beta^4}e^{-\frac{\ln^2\beta}{\mu^2}} \,,
\end{equation}
in agreement with Refs.~\cite{Cai:2019cdl,Yuan:2019wwo}. Therefore, we conclude that, as in the Dirac delta case, the chirality fraction in the IR limit scales as
\begin{equation}
    \Pi_\text{GW}(\kappa\ll 1)\propto \kappa \,.
\end{equation}

\section{Summary and conclusions}\label{sec:summary}

Understanding the universal features of the potential cosmic GW background is crucial for current and future GW detectors. We showed that finite-time, parity-violating GW sources exhibit a universal linear $f$ scaling of their chirality fraction in the low-frequency regime. This extends the universal $f^3$ scaling of Ref.~\cite{Cai:2019cdl} to parity-violating sources. We have shown that this universality holds for general quadratic Gaussian sources as well as for parity-violating trispectrum SIGWs \cite{Ragavendra:2025svk} and scalar-tensor-induced GWs \cite{Bari:2023rcw}. Note that, although we work with the collapsed shape of the trispectrum \eqref{eq:template1}, we expect the universality to hold in general, as the leading order of the IR tail of the SIGWs is generated from the collapsed limit of the trispectrum (see the discussion around Eq.~\eqref{eq:EforQs}). 

Our analysis does not depend on the generation time, as long as it occurs during the radiation-dominated early universe. However, our results apply only to models in which the chiral GW source is quadratic and active for a finite time. This excludes models for which the chiral source extends over many scales, e.g., either because it is always active during inflation or has a scale-invariant primordial spectrum \cite{Sorbo:2011rz,Caprini:2014mja,Garcia-Bellido:2016dkw,Fu:2024ipa}, and models where the source is linear, like additional spin-2 particles \cite{Bordin:2018pca,Gorji:2023ziy}. Exceptions are the models of Refs.~\cite{Obata:2016tmo,Obata:2016oym}, where the chiral GW spectrum becomes unpolarized at very low frequencies because parity violation is only effectively active after a given time, but the total GW spectrum does not decay as $f^3$ because the unpolarized source is active throughout inflation. It also does not apply to the case where the parity violation is in the gravity sector, e.g., via a gravitational Chern-Simons term \cite{PhysRevD.68.104012,Zhang:2022xmm,Feng:2023veu}, although we expect similar results if the parity-violating term is active only for a finite time. We leave a detailed study of these models and the scaling of the chirality fraction in these cases for future work.

\begin{acknowledgments}
We thank Lucas Pinol for useful discussions. This work was supported by the DFG under the Emmy-Noether program, project no. 496592360, and the JSPS KAKENHI grant no. JP24K00624.
T.C. was supported by the physics department of the École Normale Supérieure of Paris.
\end{acknowledgments}

\appendix

\section{Polarization basis \label{app:pol-basis}}

In this appendix, we explicitly write down the conventions for the polarization basis, following \cite{Domenech:2021ztg}. Namely, for the vector $\bk$ and the polarization vectors we take in general
\begin{align}\label{polvectors+x}
    \begin{split}
        \bk &= k\left(\sin\theta_k\cos\varphi_k,\sin\theta_k\sin\varphi_k,\cos\theta_k\right) \,,\\
        \be(\bk) &= \left(\cos\theta_k\cos\varphi_k,\cos\theta_k\sin\varphi_k,-\sin\theta_k\right) \,,\\
        \bar{\be}(\bk) &= \left(-\sin\varphi_k,\cos\varphi_k,0\right) \,.
    \end{split}
\end{align}
For simplicity and without loss of generality, we set $\bk$ in the z-direction, so that
\begin{align}\label{def_basis}
\begin{split}
    \bk &= k\left(0,0,1\right) \,,\\
    \be &= \left(1,0,0\right) \,,\\
    \bar{\be} &= \left(0,1,0\right) \,.
\end{split}
\end{align}
We then parametrize two general momenta, say $\bq$ and $\bl$, as 
\begin{align}
    \begin{split}
        \bl & = l\left(\sin\theta_l\cos\varphi_l,\sin\theta_l\sin\varphi_l,\cos\theta_l\right) \,,\\
        \bq & = q\left(\sin\theta_q\cos\varphi_q,\sin\theta_q\sin\varphi_q,\cos\theta_q\right) \,.
    \end{split}
\end{align}

Lastly, the polarization tensors in the linear basis are given by
\begin{align}\label{eq:poltensors_pm}
    \begin{split}
        e_{ij}^{+}(\bk) & =\frac{1}{\sqrt{2}}\left(e_{i}(\bk)e_{j}(\bk)-\bar{e}_i(\bk)\bar{e}_{j}(\bk)\right) \,,\\
        e_{ij}^{\times}(\bk) & =\frac{1}{\sqrt{2}}\left(e_{i}(\bk)\bar{e}_{j}(\bk)+\bar{e}_i(\bk)e_{j}(\bk)\right)  \,.       
    \end{split}
\end{align}
In the helicity basis, which is the one we use in the main text, we have that
\begin{align}\label{eq:poltensors_RL}
    \begin{split}
        e_i^{\pm}(\bk) & = \frac{1}{\sqrt{2}}\left(e_i(\bk)\pm i\bar{e_i}(\bk)\right) \,,\\
        e_{ij}^{\pm}(\bk) & =\frac{1}{\sqrt{2}}\left(e_{ij}^+(\bk)\pm ie_{ij}^\times(\bk)\right) \,.
    \end{split}
\end{align}
Note that
\begin{align}
\left(e_i^{s}(\bk)\right)^*e_i^{s'}(\bk)=\delta_{ss'}\quad{\rm and}\quad\left(e_{ij}^{\lambda}(\bk)\right)^*e_{ij}^{\lambda'}(\bk)=\delta_{\lambda\lambda'}\,.
\end{align}
One also has that
\begin{align}
e_{ij}^{\pm}(\bk)=e_i^{\pm}(\bk)e_j^{\pm}(\bk)\,.
\end{align}

\section{Details on the computation of SIGWs induced by the trispectrum}

In this appendix, we present details of the calculations regarding the trispectrum SIGWs. We first review the equivalence of the $f_{\rm NL}$ expansion with the $\tau_{\rm NL}$ template \eqref{eq:template1}. Then we present details of the IR limit for the Dirac delta case.

\subsection{$f_{\rm NL}$-expansion}\label{app:F_NL-expansion}
Consider small departures from Gaussianity and treat non-Gaussianities as perturbative corrections to the Gaussian variable. To this end, we write, as usual, the non-Gaussian curvature perturbation as
\begin{equation}\label{eq:F_NL-expansion}
    \mathcal{R}(\bx)=\mathcal{R}_g(\bx)+f_\text{NL}\left(\mathcal{R}^2_g(\bx)-\langle\mathcal{R}^2_g(\bx)\rangle\right) \,,
\end{equation}
where $\mathcal{R}_g$ follows Gaussian statistics and $f_\text{NL}$ is the usual expansion parameter of local-type non-Gaussianities.
The one-loop power-spectrum of $\mathcal{R}$ is then
\begin{equation}
    \mathcal{P}_\mathcal{R}(k)=\mathcal{P}_{\mathcal{R}_g}(k)+2f_\text{NL}^2\int \frac{d^3q}{(2\pi)^3}\mathcal{P}_{\mathcal{R}_g}(q)\mathcal{P}_{\mathcal{R}_g}(\left|\bk-\bq\right|) \,,
\end{equation}
where $\mathcal{P}_{\mathcal{R}_g}$ is the power spectrum of the Gaussian fluctuations. The contribution from the one-loop power-spectrum to the SIGWs is often referred to as the hybrid contribution \cite{Adshead:2021hnm,Perna:2024ehx}, which we neglect for simplicity.
Inserting the expansion Eq.~\eqref{eq:F_NL-expansion} into the definition of the trispectrum, Eq.~\eqref{eq:trispectrum_def}, one can reduce the four-point function to a sum of products of power spectra of the Gaussian curvature perturbation $\mathcal{R}_g$ with twelve permutations of the momenta.

In the following, we collect all possible permutations of momenta that appear in the $\tau_{\rm NL}$ template \eqref{eq:template1} and that also arise when inserting the $f_{\rm NL}$-expansion of the curvature perturbation, Eq.~\eqref{eq:F_NL-expansion}, into the four-point function Eq.~\eqref{eq:trispectrum_def}. They are given by
\begin{align}
    \begin{split}
        \mathscr{S} =\{&\{\bq, \bk-\bq,-\bl, -\bk+\bl\}, \{\bq, \bk-\bq, -\bk+\bl,-\bl\}, \{\bq,-\bl, \bk-\bq, -\bk+\bl\}, \{\bq,-\bl, -\bk+\bl,\bk-\bq\}\\
        & \{\bq, -\bk+\bl, \bk-\bq,-\bl\}, \{\bq, -\bk+\bl,-\bl, \bk-\bq\},\{-\bl, \bq, \bk-\bq, -\bk+\bl\}, \{-\bl, \bq, -\bk+\bl, \bk-\bq\},\\
        &\{-\bl, \bk-\bq, \bq, -\bk+\bl\}, \{-\bl, \bk-\bq, -\bk+\bl, \bq\}, \{-\bl, -\bk+\bl, \bq, \bk-\bq\}, \{-\bl, -\bk+\bl, \bk-\bq, \bq\}\} \,.
    \end{split}
    \label{eq:permutations_set}
\end{align}
These should be understood as combinations arising in a sum of the form
\begin{align}
    \mathcal{T}(\bk_1,\bk_2,\bk_3,\bk_4) & \supset  \sum_{\bk_1,\bk_2,\bk_3,\bk_4\in\mathscr{S}}P_{\cal R}(k_1)P_{\cal R}(k_3)P_{\cal R}(|\bk_1+\bk_2|)\,,
\end{align}
which directly maps to Eq.~\eqref{eq:template1}.

Following the notation of \cite{Adshead:2021hnm}, one can decompose the connected part of the trispectrum at $\mathcal{O}\left(f_{\rm NL}^2\right)$ into two pieces, labeled the ``\textit{Z}'' and ``\textit{C}'' parts, respectively, according to their Feynman-like diagrammatic structure, which we present in App.~\ref{app:feynman-diags}. One may then map the non-vanishing permutations to the \textit{Z} and \textit{C} terms as
\begin{align}\label{eq:hzc-terms}
    \begin{split}
        & \left.
        \begin{array}{ll}
            & \mathcal{T}_3 (\bq,-\bl,\bk-\bq,-\bk+\bl)=\mathcal{P}(q)\mathcal{P}(|\bk-\bq|)\mathcal{P}(|\bl-\bq|)+\bq\leftrightarrow\bl\\
            & \mathcal{T}_5 (\bq,-\bk+\bl,\bk-\bq,-\bl)=\mathcal{P}(q)\mathcal{P}(|\bk-\bq|)\mathcal{P}(|\bk-(\bq+\bl)|)+\bq\leftrightarrow\bl
        \end{array}
        \right\} \hspace{1em}(\textit{C}) \\
        & \left.
        \begin{array}{ll}
            & \mathcal{T}_4 (\bq,-\bl,-\bk+\bl,\bk-\bq)=\mathcal{P}(q)\mathcal{P}(|\bk-\bl|)\mathcal{P}(|\bl-\bq|)+\bq\leftrightarrow\bl\\
            & \mathcal{T}_6 (\bq,-\bk+\bl,-\bl,\bk-\bq)=\mathcal{P}(q)\mathcal{P}(l)\mathcal{P}(|\bk-(\bq+\bl)|)+\bq\leftrightarrow\bl
        \end{array}
        \right\} \hspace{1em}(\textit{Z}) \,,
    \end{split}
\end{align}
where we labeled the terms depending on the permutation which they arise from. Those terms are all second order in the $f_\text{NL}$ expansion.

\subsection{Feynman diagrams}\label{app:feynman-diags}

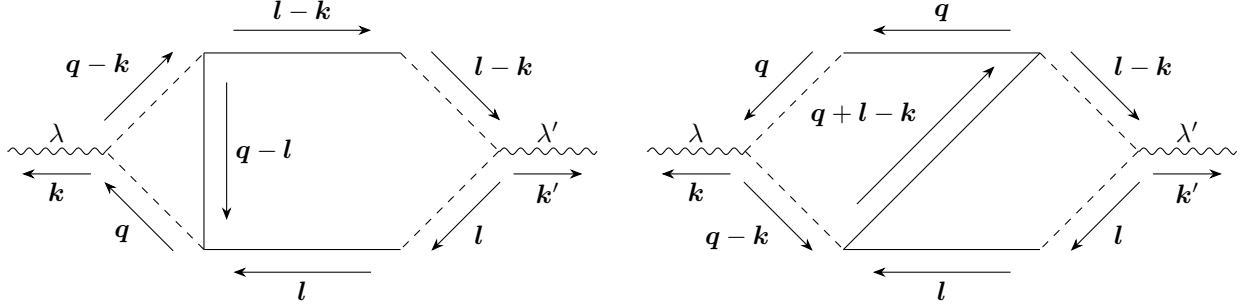
\begin{figure}
\subfloat{
\begin{tikzpicture}[scale=1.3]
\begin{feynman}
    \vertex (a) at (0,1);
    \vertex (b) at (1,1);
    \vertex (c) at (2,2);
    \vertex (d) at (2,0);
    \vertex (e) at (4,2);
    \vertex (g) at (4,0);
    \vertex (f) at (5,1);
    \vertex (h) at (6,1);
    \diagram* {
        (a) -- [photon, edge label=\(\lambda\), reversed momentum'=\(\bk\)] (b),
                  (b) -- [scalar, momentum=\(\bq-\bk\)] (c),
                  (c) -- [momentum=\(\bq-\bl\)] (d),
                  (d) -- [scalar, momentum=\(\bq\)] (b),
                  (c) -- [momentum=\(\bl-\bk\)] (e),
                  (e) -- [scalar, momentum=\(\bl-\bk\)] (f),
                  (f) -- [scalar, momentum=\(\bl\)] (g),
                  (g) -- [momentum=\(\bl\)] (d),
                  (f) -- [photon, edge label=\(\lambda'\), momentum'=\(\bk'\)] (h)
                };
\end{feynman}
\end{tikzpicture}
}
\hspace{0.3cm}
\subfloat{
\begin{tikzpicture}[scale=1.3]
\begin{feynman}
    \vertex (a) at (0,1);
    \vertex (b) at (1,1);
    \vertex (c) at (2,2);
    \vertex (d) at (2,0);
    \vertex (e) at (4,2);
    \vertex (g) at (4,0);
    \vertex (f) at (5,1);
    \vertex (h) at (6,1);
    \diagram* {
    (a) -- [photon, edge label=\(\lambda\), reversed momentum'=\(\bk\)] (b),
    (b) -- [scalar, reversed momentum=\(\bq\)] (c),
    (b) -- [scalar, momentum'=\(\bq-\bk\)] (d),
    (c) -- [opacity=0] (d),
    (c) -- [reversed momentum=\(\bq\)] (e),
    (g) -- [momentum=\(\bl\)] (d),
    (d) -- [momentum=\(\bq+\bl-\bk\)] (e),
    (e) -- [scalar, momentum=\(\bl-\bk\)] (f),
    (f) -- [scalar, momentum=\(\bl\)] (g),
    (f) -- [photon, edge label=\(\lambda'\), momentum'=\(\bk'\)] (h),
    (e) -- [opacity=0] (g),
    (c) -- [opacity=0] (e)
    };
\end{feynman}
\end{tikzpicture}
}
\caption{\textit{C} (left) and \textit{Z} (right) diagrams contributing to the GW power spectrum at $\mathcal{O}\left(f_\text{NL}^2\right)$. Diagrams and caption reproduced from \cite{Adshead:2021hnm}.}
\label{fig:diagrams-regular}
\end{figure}

In order to map the permutations arising in the trispectrum used in \cite{Ragavendra:2025svk}, we make use of the diagrams proposed in \cite{Adshead:2021hnm}, which we display in \cref{fig:diagrams-regular}.
The permutations $\{4,5,9,10\}$ give the same diagrams as the permutations $\{3,6,7,8\}$ shown in Fig.~\ref{fig:diagrams-regular}, but exchanging $\bl \longleftrightarrow \bl -\bk$. However, one always has the freedom to redefine the internal momenta and, furthermore, the integration plane is symmetric under such an exchange. Thus, these permutations give exactly the same contribution.

\subsection{Trispectrum SIGWs from a Dirac delta spectrum}\label{app:dirac-delta-energy-density}

Here we present the details of the calculations that we skipped in Sec.~\ref{sec:diracdeltaSIGWs}. Before we start, let us emphasize that when computing the Dirac delta case for the trispectrum in Eq.~\eqref{eq:omegaSIGWs}, one has to be careful when dealing with the variables $u_l$ and $u_q$. To see this, recall that
\begin{equation}
    |1-v_{p}|<u_{p}<1+v_{p}~,\hspace{1em}p=q,l\,.
\end{equation}
If $u_{p}$ is set to $1/\kappa$ through the action of a Dirac delta function, this equality tells us that its corresponding $v_{p}$ must asymptotically go to infinity in the IR limit. To avoid potential confusion, we note that the integration domain is symmetric under the exchange of $v_{p}$ and $u_{p}$. One can then integrate conversely $u_{p}$ from $0$ to $\infty$ and $v_{p}$ between $|1-u_{p}|$ and $1+u_{p}$. As a consequence, the integration domain for $v_{p}$ is set to $\big[|1-\frac{1}{\kappa}|,1+\frac{1}{\kappa}\big]$. One can then make the approximation
\begin{align}
    \int_{|1-\frac{1}{\kappa}|}^{1+\frac{1}{\kappa}}f(u)du &= \Theta(1-\kappa)\int_{\frac{1}{\kappa}-1}^{\frac{1}{\kappa}+1}f(u)du\\
    & \approx \Theta(1-\kappa)\times 2f\left(\frac{1}{\kappa}\right) \,.
\end{align}
Furthermore, the third spectrum, which depends on two momenta in Eq.~\eqref{eq:template1}, also requires care. Indeed, one is left with five different types of sum, and some of them hold a term proportional to $\delta(k-k^*)$. These terms are each proportional to $\cos(2\psi)$ in $\Omega^\text{even}_\text{GW}$ and $\sin(2\psi)\sin\psi$ in $\Omega^\text{odd}_\text{GW}$, with the trigonometric functions depending on $2\psi$ arising from $Q^\pm(k,v)$ and $\sin\psi$ coming from the scalar triple product in the trispectrum. The integration over $\psi$ thus yields a vanishing contribution from permutations ($\{1,2,11,12\}$).

One may then perform a change of variable to compute the remaining terms. In general, they are proportional to
\begin{equation}
    \delta(h(\cos\psi)) \,, \quad \text{where} \quad h(x)=A\sqrt{B-\sqrt{C}x} \,,
\end{equation}
with $A$ being a function of $\kappa$ and $B,\, C$, polynomials of the remaining variables. Our previous discussion ensures that the square roots give real numbers. One may thus set $x=\cos\psi$ in the integral and use the composition property of the Dirac delta function, as $\delta(f(x))=\frac{1}{|f'(x)|}\delta(x-\tilde{x})$, where $\tilde{x}\in[-1,1]$ and $|f'(x)|$ does not vanish over the interval $[-1,1]$.
Integration over $x$ leads thus to
\begin{equation}\label{eq:omega-dirac-calculation}
    \Omega^\text{even}_\text{GW}=\sum_{i=1}^{12}\iint_0^\infty dv_qdv_l\int_{|1-v_q|}^{1+v_q}du_q\int_{|1-v_l|}^{1+v_l}du_l~\mathcal{P}(|\mathscr{S}_{i,1}|)\mathcal{P}(|\mathscr{S}_{i,2}|)\mathcal{K}(v_q,u_q,v_l,u_l)\frac{2\tilde{x}^2-1}{\sqrt{1-\tilde{x}^2}}\Theta(1-|\tilde{x}|) \,,
\end{equation}
where we introduced $\mathcal{K}$ as the kernel of the integral and $\tilde{x}$ is given by
\begin{equation}
    \tilde{x}=\left\{
    \begin{array}{ll}
         \frac{-2+\kappa^2}{-4+\kappa^2} & \quad\text{for \textit{Z} terms} \\
         \frac{(-1+s^2+2t+t^2)\kappa}{2\sqrt{(1-s^2)t(2+t)(4-\kappa^2)}} & \quad\text{for \textit{C} terms}
    \end{array}
    \right.\,,
\end{equation}
where we also introduced the variables
\begin{align}
    & s=u_{p}-v_{p} \,,\\
    & t=u_{p}+v_{p}-1 \,,
\end{align}
where $p$ refers to the momentum remaining after the integration over both power spectra in Eq.~\eqref{eq:omega-dirac-calculation}.
While for \textit{Z} terms $|\tilde{x}|$ is always smaller than 1, this condition imposes that the upper boundary of the remaining integrals of the \textit{C} terms be $4/\kappa^2$. More precisely, for the third term of the permutations, one has
\begin{equation}
    \Omega_{\text{GW}}^{\text{even}}[3]=\int_0^{\infty}dt\int_{-1}^1ds~\mathcal{K}(s,t,\kappa)\frac{2\tilde{x}^2-1}{\sqrt{1-\tilde{x}^2}}\Theta(1-|\tilde{x}|) \,.
\end{equation}
The contribution of $s$ to the integral is negligible with respect to that of $t$, and one may rewrite this as
\begin{align}
    \Omega_{\text{GW}}^{\text{even}}[3] &= 2\int_0^{\infty}dt~\mathcal{K}(t,\kappa)\Theta(1-|\tilde{x}|) \approx 2\int_0^{4/\kappa}dt~\mathcal{K}(t,\kappa) \approx 2\,\mathcal{K}\left(t=\frac{4}{\kappa},\kappa\right) \,,
\end{align}
by taking the leading order in $\kappa$ and $1/t$.

The calculations for the odd trispectrum \eqref{eq:Todd} follow exactly the same procedure but are multiplied by the factor $\beta$ in Eq.~\eqref{eq:beta}, which leads to an additional factor $\kappa$ in the IR limit. Indeed, if we consider, for example,  $\bk_1=\bq$, $\bk_2=\bk-\bq$ and $\bk_3=\bl$, we have that
\begin{align}
    \beta(\bq,\bk-\bq,-\bl)=\frac{-\bq\cdot((\bk-\bq)\times\bl)}{ql|\bk-\bq|}=\frac{\sin\theta_q\sin\theta_l\sin(\varphi_q-\varphi_l)}{\sqrt{1+v_q^2-2v_q\cos\theta_q}}\,.
\end{align}
One Dirac delta sets $v_q=1/\kappa$, which leads to $\beta\propto \kappa$ in the IR limit, that is, in the $\kappa\ll1$ regime.

\section{Second template for the trispectrum}\label{app:template_2}

We also considered the second template studied in Ref.~\cite{Ragavendra:2025svk} for completeness. It is given by
\begin{align}\label{template2}
    \mathcal{T}_{\text{even}} =  &\, 2\tau_{\text{NL}}\mathcal{F}(k_1,k_2,k_3,k_4)P(k_1)P(k_3)P(|\bk_1+\bk_2|)+\text{11 permutations}\,,\\
    \mathcal{T}_{\text{odd}}= &\, i\tilde{\tau}_{\text{NL}}\mathcal{F}(k_1,k_2,k_3,k_4)\beta(\widehat{\bk_1+\bk_2},\hat{\bk_1},\hat{\bk_3})P(k_1)P(k_3)P(|\bk_1+\bk_2|)
     +\text{23 permutations}\,,
\end{align}
with
\begin{equation}
    \mathcal{F}(k_1,k_2,k_3,k_4)=1+\frac{\gamma}{\mu\sqrt{2\pi}}e^{-\frac{1}{2\sigma^2}\ln^2\frac{k_1+k_2+k_3+k_4}{4k^*}} \,.
\end{equation}
Following the calculations discussed in Sec.~\ref{sec:Trispectrum} for the first template, we obtain, for a Dirac delta peak,
\begin{align}
        \Omega_{\text{GW}}^{\text{even}}[Z] &\propto -\tau_{\text{NL}}\mathcal{A}^3_\mathcal{R}\left(\sqrt{2\pi}+\frac{\gamma}{\sigma}\right)\kappa^3\ln^2\kappa \,,\\
        \Omega_{\text{GW}}^{\text{even}}[C] &\propto \tau_{\text{NL}}\mathcal{A}^3_\mathcal{R}\kappa^2\ln^2\kappa\int_ 0^1 dx~\frac{x(2x^2-1)}{(1+8x^2)^{3/2}\sqrt{1-x^2}}\left[2\pi+\sqrt{2\pi}\frac{\gamma}{\sigma}e^{-\frac{\ln^2(x+\frac{1}{2})}{2\sigma^2}}\right] \,,
\end{align}
for the even part, and
\begin{align}
        \Omega_{\text{GW}}^{\text{odd}}[Z] &\propto \tilde{\tau}_{\text{NL}}\mathcal{A}^3_\mathcal{R}\left(\sqrt{2\pi}+\frac{\gamma}{\sigma}\right)\kappa^4\ln^2\kappa \,, \\   
        \Omega_{\text{GW}}^{\text{odd}}[C] &\sim \tilde{\tau}_{\text{NL}}\mathcal{A}^3_\mathcal{R}\kappa^3\ln^2\kappa\int_ 0^1 dx~\frac{x^2\sqrt{1-x^2}}{(1+8x^2)^{3/2}}\left[1+\sqrt{\frac{\pi}{2}}\frac{\gamma}{\sigma}e^{-\frac{\ln^2(x+\frac{1}{2})}{2\sigma^2}}\right] \,,
\end{align}
for the odd part, respectively.

We find similar results for the log-normal spectrum, which exhibits the same symmetries as in Eq.~\eqref{eq:ln-energy-density}, namely
\begin{align}
         \omega^{\text{even},(3)}_\text{GW} \sim & \,18\sqrt{8}\pi^{5/2}\tau_{\text{NL}}\mathcal{A}^3_\mathcal{R}\kappa^3\iint_0^\infty\frac{d\alpha}{\alpha}\frac{d\beta}{\beta}\ln\frac{\alpha}{\kappa}\ln\frac{\beta}{\kappa}\left(\frac{\alpha}{\beta}\right)^3\left(\sqrt{2\pi}+\frac{\gamma}{\sigma}e^{-\frac{\ln^2((\alpha+\beta)/2)}{2\sigma^2}}\right)\frac{1}{\mu}e^{-\frac{\ln^2\beta}{2\mu^2}} \nonumber\\
         & \times\int_0^{2\pi}d\psi~\frac{1}{\mu}e^{-\frac{\ln^2\sqrt{f(\psi)}}{2\mu^2}}\frac{\cos(2\psi)}{f^{3/2}(\psi)} \,,
\end{align}
and
\begin{align}
        \omega^{\text{odd},(3)}_\text{GW} \sim &\, 18\pi^{5/2}\tilde{\tau}_{\text{NL}}\mathcal{A}^3_\mathcal{R}\kappa^4\iint_0^\infty\frac{d\alpha}{\alpha}\frac{d\beta}{\beta}\left(\ln^2\kappa+\ln\alpha\ln\beta\right)\frac{\alpha^3}{\beta^4}\left(\sqrt{2\pi}+\frac{\gamma}{\sigma}e^{-\frac{\ln^2((\alpha+\beta)/2)}{2\sigma^2}}\right)\frac{1}{\mu}e^{-\frac{\ln^2\beta}{2\mu^2}} \nonumber\\
         & \times\int_0^{2\pi}d\psi~\frac{1}{\mu}e^{-\frac{\ln^2\sqrt{f(\psi)}}{2\mu^2}}\frac{\sin\psi\sin(2\psi)}{f^{3/2}(\psi)}\ \,.
\end{align}
Hence, we obtain a similar scaling to that found for the first template \eqref{eq:template1}.

\bibliography{biblio}

\end{document}